# A novel scheme for hundred-hertz linewidth measurements with self-heterodyne method

Yu Peng

*School of Physics, Beijing Institute of Technology, Beijing, 100081, P. R. China*

**Abstract**

We propose a novel scheme for accurately determining hundred-hertz linewidth by delayed self-heterodyne method in which delay time far less than the laser's coherence time. That exceeds the former understanding as to the delayed self-heterodyne technique which requiring a prohibitively long fiber. The self-heterodyne autocorrelation function and power spectrum are evaluated and by numerical analysis we ensure that -3dB of power spectrum is applied to the self-heterodyne linewidth measurements. For laser linewidth less than 100 Hz, linewidth can be measured directly by 10 km fiber, and in more general case linewidth can be deduced from -20 dB or -40 dB of the fitting Lorentzian curve.

**1. Introduction**

Narrow linewidth lasers are highly desirable for applications such as optical atomic clock work [1-4], gravitational wave detection [5-7], cavity quantum electrodynamics [8], quantum optomechanics [9-10] and precision tests of relativity [11]. Accurate measurements of these narrowed spectral linewidth are required to characterize system performance. Two methods have been used for laser linewidth determination. The most common method is using another linewidth laser to beatnote the signal and measures these narrowed linewidth. The other one is delayed self-heterodyne technique. This technique first proposed by Okoshi *et al.* [12] offers the highest resolution, but requires that the delay time is much longer than the laser's coherence time.

Using this method, however, it is practically difficult to measure linewidth in the kHz and Hertz range where requiring a prohibitively long fiber since reliable measurements in the presence of 1/f noise require delays 5–6 times longer than the coherence length of the laser [13]. Long delays are required for these measurements to provide adequate resolution.

Here we propose a new scheme for accurately determining hundred-hertz linewidth by delayed self-heterodyne method in which delay time significantly less than the laser's coherence time. That is beyond former understanding about self-heterodyne technique. From numerical analysis we ensure that -3dB of power spectrum is applied to the self-heterodyne linewidth measurements with just 10 km fiber.

**2. Principle and analysis**

The diagram is illustrated in Fig. 1. The laser produced a narrow-line source with which we could evaluate this self-heterodyne technique. The output from this narrow-linewidth laser was directed through an optical isolator into the interferometer. The optical isolator is used to avoid unwanted feedback from the fiber. Immediately following the isolator, the beam was split into two paths by a 50/50 beam splitter. The remainder of the beam was delayed by single-mode optical fiber. The other beam was directed through an acousto-optic modulator operating at 80 MHz. The two beams were recombined by a beam splitter and then detected by a Si avalanche photodiode (APD).

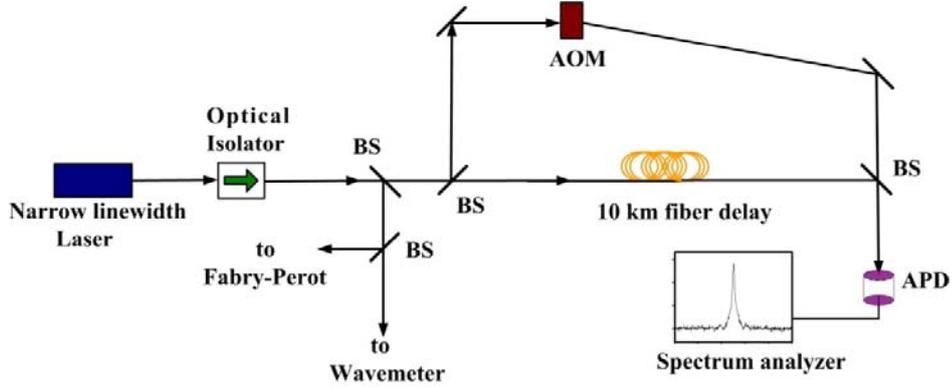

**FIG.1.** (Color online) Diagram of the self-heterodyne measurement for a narrow-linewidth laser. BS: beam splitter, AOM: acousto-optic modulator, APD: Si avalanche photodiode, FP: Fabry-Perot interferometer.

Briefly reviewing the interferometric theory, the total electric field incident on the detector $E_d(t)$ is found by summing the electric field from each arm of the interferometer:

$$E_d(t) = \sqrt{P_0}\cos(\omega_0 t + \phi(t)) + \sqrt{P_0}\cos((\omega_0 + \Omega)\bullet(t-\tau) + \phi(t-\tau)) \qquad (1)$$

Where $P_0$ is the laser's output power, $\omega_0$ is the laser frequency, $\tau$ is the time delay of one path with respect to the other path, and $\Omega$ is the offset frequency. The total detected intensity is then given by

$$I_d(t) = P_0 + P_0\cos[-\Omega t + (\omega_0 + \Omega)\tau + \phi(t) - \phi(t-\tau)] \qquad (2)$$

The resulting autocorrelation function $\gamma(\delta\tau,\tau)$ is then found to be

$$\gamma(\delta\tau,\tau) = \frac{1}{2}P_0^2 \cos(\Omega\delta\tau) < \cos(\Delta\phi(\delta\tau,0) - \Delta\phi(\delta\tau-\tau,\tau)) > \qquad (3)$$

Where $\Delta\phi(t_1,t_2) = \phi(t'+t_1) - \phi(t'+t_2)$ is the phase difference between times $t_1$ and $t_2$. The relationship $<\cos x> = e^{-<x^2>/2}$ can be used to simplify the expression for $\gamma(\delta\tau,\tau)$. In addition, the phase correlation of the recombined beams can be written as

$$<[\phi(t)-\phi(t-\tau)]^2> = \frac{\tau}{\tau_c} \qquad (4)$$

Where $\tau_c$ is the coherence time, then, with further manipulation, the autocorrelation function may ultimately be expressed as

$$\gamma(\delta\tau,\tau) = \frac{1}{2}P_0^2 e^{-\tau/\tau_c} \cos(\Omega\delta\tau) \bullet \exp[-\frac{1}{\tau_c}(\delta\tau-\tau)] \qquad (5)$$

Using the Wiener-Khintchine theorem and performing the integration, we arrive at the following expression for power spectral density:

$$S(\omega,\tau) = \frac{\frac{1}{2}P_0^2 \tau_c}{1+(\omega\pm\Omega)^2 \tau_c^2}\{1-e^{-\tau/\tau_c}[\cos[(\omega\pm\Omega)\tau]+\frac{\sin[(\omega\pm\Omega)\tau]}{(\omega\pm\Omega)\tau_c}]\} \\ + \frac{1}{2}P_0^2 \pi e^{-\tau/\tau_c}\delta(\omega\pm\Omega) \qquad (6)$$

The performance of $\frac{1}{2}P_0^2\tau_c/(1+\omega^2\tau_c^2)$ is Lorentzian line, and performance of $1-e^{-\tau/\tau_c}[\cos(\omega\tau)+\sin(\omega\tau)/(\omega\tau_c)]$ is periodically fluctuations superimposed on the Lorentzian line. The fluctuations are mainly caused by the item of $\cos(\omega\tau)$. When the fiber delay line in infinitely long, the power spectral density can be written in standard Lorentzian line:

$$S(\omega,\tau) = \frac{P_0^2}{4\pi}\frac{\Delta v}{\Delta v^2 + v^2} \qquad (7)$$

Where $\Delta v = 1/(2\pi\tau_c)$ is the intrinsic linewidth of laser, and $v = \omega/(2\pi)$ is the measured frequency. This equation indicates that as the delay time increases, the signal strength shifts from the delta function peak to the modified Lorentzian pedestal until the power spectrum becomes strictly Lorentzian. This true Lorentzian corresponds to the delay time when the phases of the optical field have become totally decorrelated.

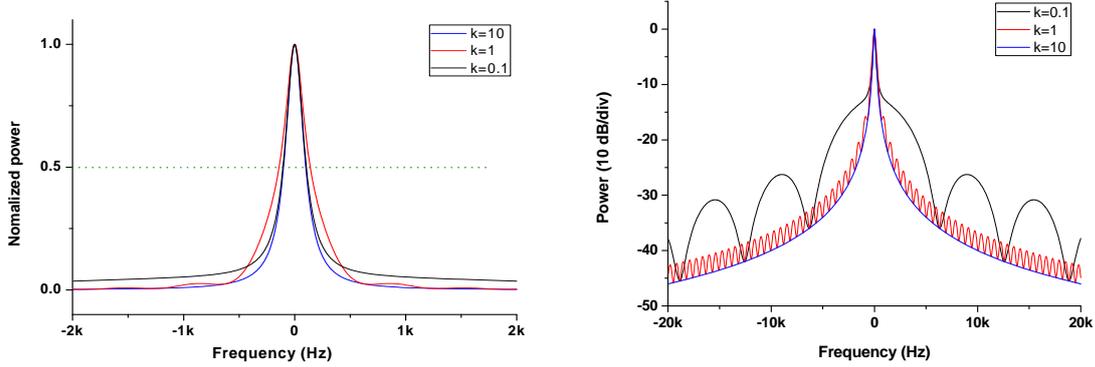

**FIG.2.** (Color online) Power spectrum comparisons of self-heterodyne signal with different lengths of delay line, $k = 0.1$ (black), $k = 1$ (red), and $k = 10$ (blue). (a) Signal in linear scale. (b) Signal in logarithmic scale.

This behavior is illustrated in Fig. 2. In our scheme we assume that the true linewidth of laser is 100 Hz, and correspondingly coherence time of laser is 1.59 ms. The power spectrum was observed with three different lengths of fiber, 31.8 km, 318 km, and 3183 km respectively. These lengths correspond to delay times of approximately 159 μs, 1.59 ms, and 15.9 ms, with the value of ratio $k = \tau/\tau_c$, 0.1, 1 and 10 respectively. All measurements were taken with the laser operating at the same average output power level. Fig. 2(a) shows the power spectral density in a linear scale. And Fig. 2(b) show that as k increase, the sideband ripples disappear. In all of our computations, the predicted laser linewidth is considered to be half of -3 dB bandwidth (FWHM).

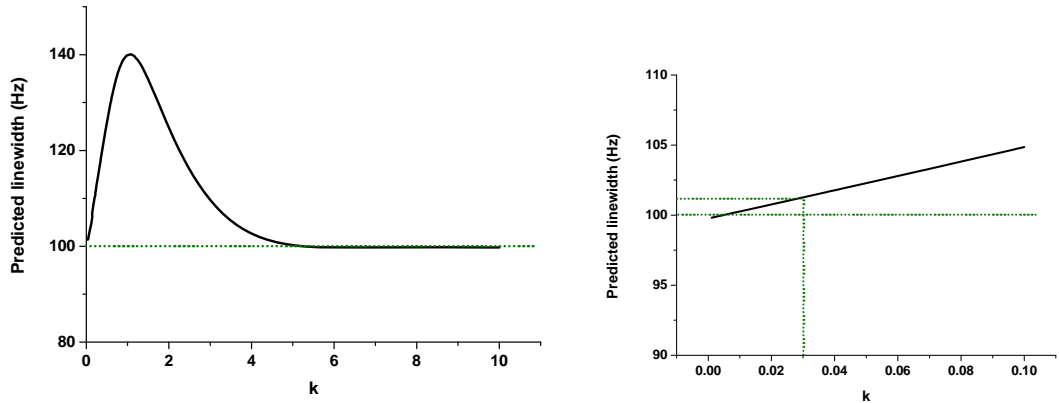

**FIG. 3.** (Color online) Relationship between predicted linewidth and k. (a) k<10 (b) k<0.1

Our computations have shown that in order to measure the width of the pedestal portion of the beat signal, the delay time must be more than 5 times the laser's coherence time, shown in Fig. 3(a). The theoretically predicted behavior is that the sideband ripples as k was increased towards 1, and then disappears as k exceeded 5. It implies that a fiber optic delay of more than 1590 km would be required to enable the true measurement of the 100-Hz linewidth. Furthermore by analyzing the relationships between the predicted linewidth (half of measured the -3 dB bandwidth) and the ratio k, shown in Figure 3(b), we found that we can measure the linewidth with k<0.1 area which means short fiber delay although there are some deviations. With 10 km fiber delay the linewidth is theoretically predicted to be 101.2 Hz which include the deviation of 1.2 Hz, and the accuracy achieve 98.8 %.

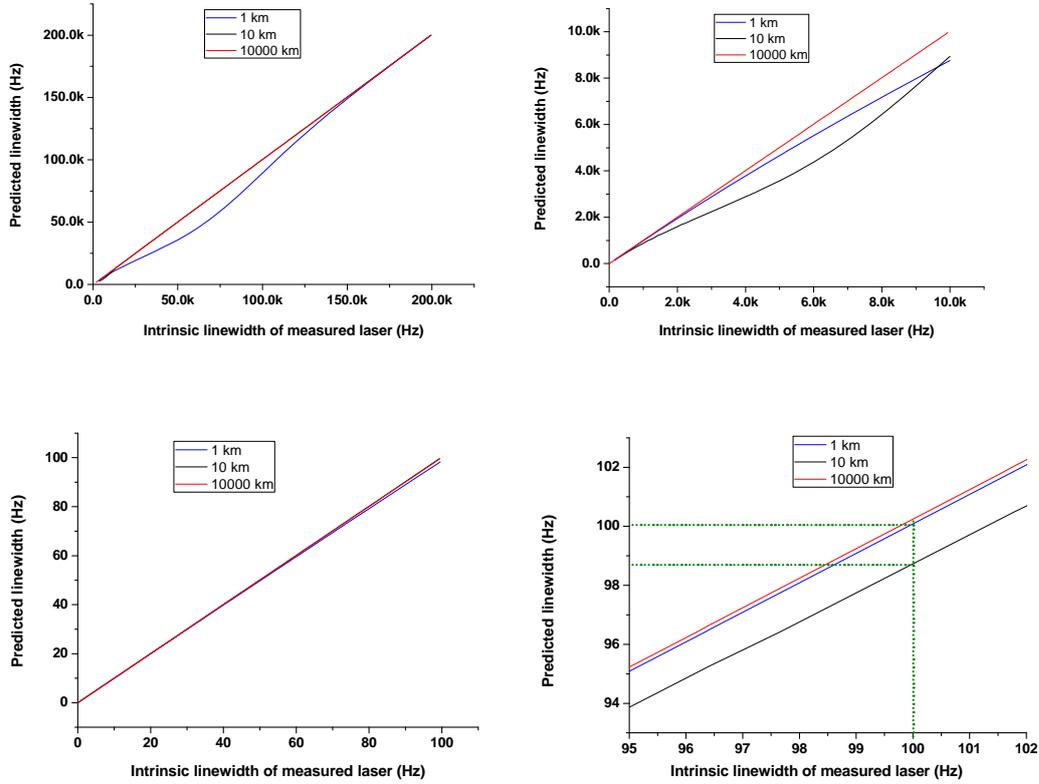

**FIG. 4.** (Color online) Calculated linewidth versus intrinsic linewidth of laser with fiber delay, 1 km (blue), 10 km (black) and 10000 km (red). (a) Span: 200 kHz. (b) Span: 10 kHz. (c) Span: 100 Hz. (d) Inset around 100 Hz.

For further analysis, we show the predicted linewidth with fiber length of 1 km, 10 km and 10000 km (infinite length of delay line) respectively, corresponding to delay times of 5 μs, 50 μs, 50 ms separately, in Figure 4. We demonstrate three of Fourier frequency, 200 kHz, 10 kHz and 100 Hz respectively. In the range of 10 kHz to 200 kHz, the deviations caused by 1 km fiber between predicted linewidth and laser intrinsic linewidth is more remarkable, shown in Fig. 4(a). And in the range of 1 kHz to 10 kHz, the deviation caused by 10 km fiber is significant in Fig. 4(b). But the predicted linewidth and intrinsic linewidth are approximately equal to each other in the area less than 100 Hz in Fig.4(c). And the predicted linewidth is 98.8 Hz which agrees with infinite length of delay line well, shown in Fig.4(d).

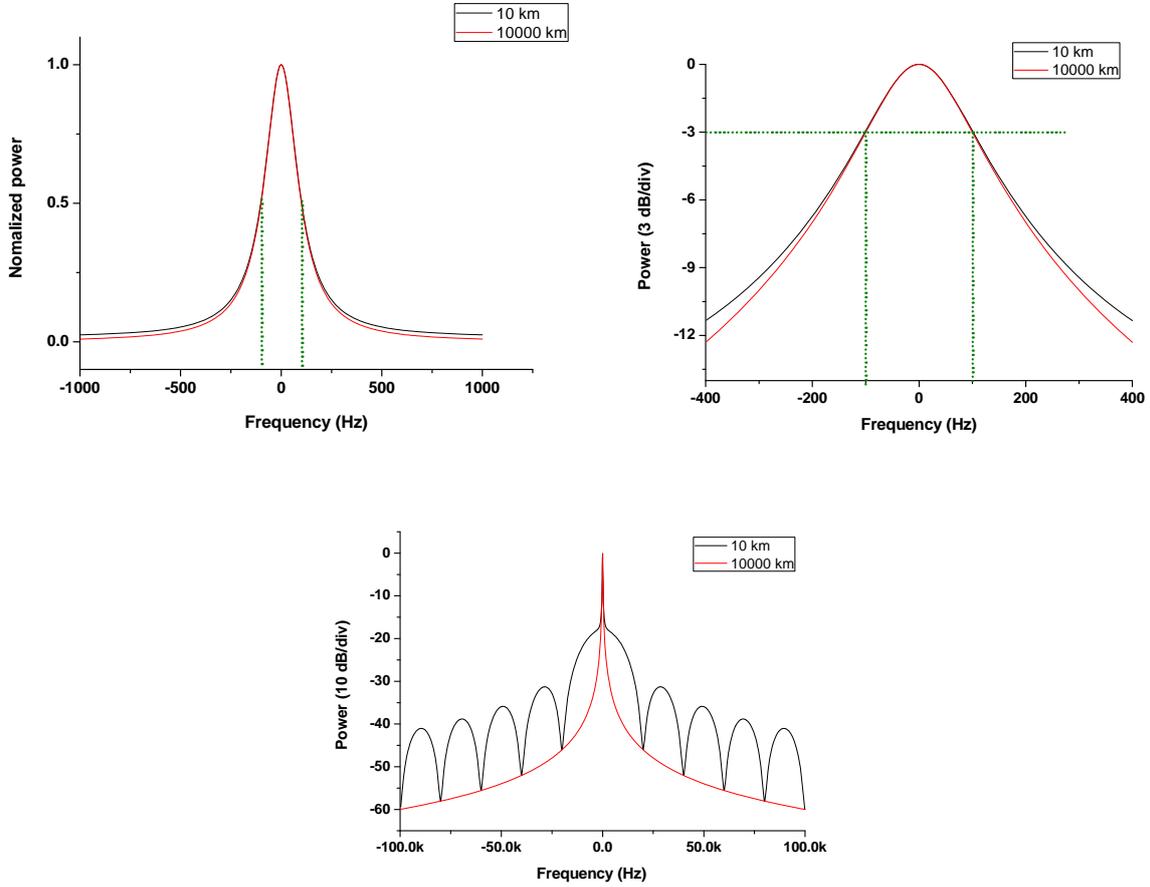

**FIG.5.** (Color online) Power spectrum comparisons with different lengths of delay line, 10 km and 10000 km, correspondingly $k$ =0.031 (black) and $k$ =31 (red). (a) Power spectral density signal in linear scale. (b) Power spectral density signal in logarithmic scale.

The single-mode optical fiber used as a delay line had a transmission loss of approximately 3 dB/km in this wavelength region of 698 nm. Fig. 5 shows the power spectrum with two lengths of delay line, 10 km and 10000 km respectively, corresponding to the value of k, 0.031 and 31. The FWHM is obtained by analyzing the power spectral density in a linear scale, shown in Fig. 5(a). The FWHM are 202.68 Hz and 199.52 Hz, which corresponds to the predicted linewidth of

101.34 Hz, and 99.76 Hz.

Moreover, the power spectrum of self-heterodyne signal in logarithmic scale is shown in Fig. 5(b). The deduced laser linewidth from the width of the line shape -3 dB from the peak, 99.84 Hz for 10 km delay, agree with 101.12 Hz for 10000 km delay used for the measurement. These two values, we think, agree with each other well and there is no remarkable difference in results. Both accuracies achieve more than 98 %. Therefore we conclude that measuring narrow linewidth with short fiber delay by the power spectrum of -3 dB directly is valid.

But for the line shape -20 dB where the effects of the 1 /f noise are great, the presented results show that the measured FWHM can depend substantially upon the delay, shown in Fig. 5(c). In the case of 10000 km, the value of -40 dB is 9999.5 Hz, and value of -20 dB is 994.98 Hz, which agree with intrinsic linewidth of measured laser well. And for 10 km of delay line, the value of -20dB is 7265.7 Hz and the value of -40dB is 37026 Hz, which is modified by 1 /f noise dramatically.

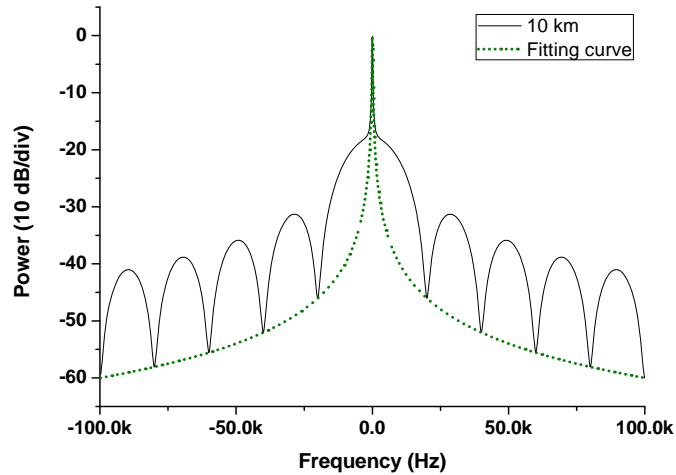

**FIG.6.** (Color online) Fitting the power spectrum of self-heterodyne signal by a Lorentzian curve in logarithmic scale (green).

For more general case, especially laser with linewidth more than 100 Hz, we can estimate the predicted linewidth with 10 km fiber delay too. By fitting the power spectrum of self-heterodyne signal by a Lorentzian curve along the bottom of signal, the green line as shown in Fig. 6, the laser linewidth was simply deduced from -20dB or -40dB of fitting curve. The figure shows -20 dB linewidth of the fitting curve is 996.72 Hz, therefore we deduce that -3dB linewidth is 99.67 Hz with accuracy of 99.6 %.

## 3. Conclusion

In summary, a novel scheme of self-heterodyne technique is proposed for hundred-hertz linewidth measurements, in which delay time far less than the laser's coherence time. We believe that this technique may provide a viable way of measuring the linewidth of a stable narrowed laser based on the analysis of these results. For laser with linewidth less than 100 Hz, linewidth can be predicted to be measured directly by 10 km fiber. In more general case fitting method can be used and the linewidth can be deduced from -20dB or -40dB of fitting curve. These results exceed the former understanding as to the delayed self-heterodyne technique and it can be applied in roughly analyzing narrow linewidth in the future.

The accuracy of all these results achieved over 98%, but in real circumstance, the results must be influenced due to fiber sensitivity to temperature, pressure, and tension etc. The real circumstance temperature was kept at 25 ℃. And the change of strain and temperature of the fiber are estimated to be within 100 μstrain and an 2 ℃, respectively. The thermal expansion coefficient of single mode communication optical fiber (SMF-28) is $\alpha_T = 5.5 \times 10^{-7} / {}^o C$ [14]. According to $\frac{\Delta L}{L} = \alpha_T \Delta T$, the inaccuracy linewidth caused by the change of fiber length due to circumstance temperature, we estimate, less than 0.03%. The change of effective refractive index of the fiber core is $\Delta n$. Assuming that the strain- and thermally-induced perturbations are linear, in response to a strain change $\Delta \varepsilon$ and temperature change $\Delta T$. The change due to strain and temperature, $\Delta n$ can be expressed in the form $\Delta n_\varepsilon = C_\varepsilon \Delta \varepsilon$, and $\Delta n_T = C_T \Delta T$ separately. The strain and temperature coefficients of fiber optic refractive index were $C_\varepsilon = -0.36 \times 10^{-6} / \mu\varepsilon$ and $C_T = 1.0 \times 10^{-5} / {}^oC$ for 1300 nm wavelength, respectively [14]. And the change of $\Delta n$ cause the inaccuracy linewidth which we estimate is about 0.05%. This assumes that the strain and thermal response are essentially independent, i.e. the related strain-temperature cross-term is negligible, which has already been found to apply well for small perturbations.

**Acknowledgments**

The author thanks Dr. Erjun Zang (National Institute of Metrology, China) for his useful discussions. This work is supported by Basic Research Foundation from Beijing Institute of Technology (Grant No. 20121842004).